\documentclass[%
 reprint,%
 amssymb, amsmath,%
 aip,cha,%
]{revtex4-1}

\usepackage{docs}%
\usepackage{bm}%
\usepackage{graphicx}%
\usepackage{subfigure}%
\usepackage[colorlinks=true,linkcolor=blue]{hyperref}%
\expandafter\ifx\csname package@font\endcsname\relax\else
 \expandafter\expandafter
 \expandafter\usepackage
 \expandafter\expandafter
 \expandafter{\csname package@font\endcsname}%
\fi
\begin{document}

\title{Graphene mode locked, wavelength-tunable, dissipative soliton fiber laser}%

\author{Han Zhang,$^{1}$ Dingyuan Tang,$^{1,*}$ R. J. Knize,,$^2$ Luming Zhao,$^{1}$ Qiaoliang Bao,$^3$ and Kian Ping Loh ,$^3$}%

\address{
$^1$School of Electrical and Electronic Engineering, Nanyang Technological University, Singapore 639798
\\
$^2$Department of Physics, United States Air Force Academy, Colorado 80840, United States of America \\
$^3$Department of Chemistry, National University of Singapore, Singapore 117543 \\
$^*$Corresponding author: edytang@ntu.edu.sg, zhanghanchn@htomail.com\\
}
\email{http://www3.ntu.edu.sg/home2006/zhan0174/}

\date{February 2010}%
\revised{\today}%
\maketitle

\tableofcontents

\section{Introduction}

Atomic layer graphene possesses wavelength-insensitive ultrafast saturable absorption, which can be exploited as a "full-band" mode locker. Taking advantage of the wide band saturable absorption of the graphene, we demonstrate experimentally that wide range (1570 nm - 1600nm) continuous wavelength tunable dissipative solitons could be formed in an erbium doped fiber laser mode locked with few atomic layer graphene.

\subsection{Nanotube Mode-Locked Fiber Laser}

Recently, passive mode locking of fiber lasers with single walled carbon nanotubes (SWCNTs) has attracted considerable attention \cite{no1,no2,no3,no4,no5,no6,no7}.  It has been shown that SWCNT mode lockers have the advantages such as intrinsically ultrafast recovery time, large saturable absorption, easy to fabricate, and low cost. In particular, as SWCNTs are direct-bandgap materials with a gap that depends on the nano-tubes' diameter and chirality, through mixing SWCNTs with different diameters, a broadband saturable absorption mode locker could be made. A wideband wavelength tunable erbium-doped fiber laser mode locked with SWCNTs was experimentally demonstrated \cite{no2}.

\subsection{Drawback of Nanotube Mode-Locker}

However, the broadband SWCNT mode locker suffers intrinsic drawbacks: SWCNTs with a certain diameter only contribute to the saturable absorption of a particular wavelength of light, and SWCNTs tend to form bundles that finish up as scattering sites. Therefore, coexistence of SWCNTs with different diameters introduces extra linear losses to the mode locker, making mode locking of a laser difficult to achieve. In this letter we show that these drawbacks could be circumvented if graphene is used as a broadband saturable absorber. Implementing graphene mode locking in a specially designed erbium-doped fiber laser, we demonstrated the first wide range (1570nm- 1600nm) wavelength tunable dissipative soliton fiber laser.

\subsection{Soliton Fiber Laser}

Soliton operation of mode locked fiber lasers has been extensively investigated previously. Conventionally, the study on soliton operation of fiber lasers was focused on lasers with anomalous cavity dispersion, where a soliton is formed due to the natural balance between the anomalous cavity dispersion and fiber nonlinear optical Kerr effect, and the dynamics of the formed solitons is governed by the nonlinear Schrodinger equation. Recently, it was further shown both experimentally and theoretically that a soliton could even be formed in fiber lasers with large normal cavity dispersion \cite{no8,no9}. Formation of solitons in the normal dispersion fiber lasers is a result of the mutual nonlinear interaction among the normal cavity dispersion, fiber Kerr nonlinearity, and the effective laser gain bandwidth filtering \cite{no8)}, and the dynamics of the formed solitons is governed by the complex Ginzburg-Landau equation (GLE). A soliton whose dynamics is governed by the GLE is also known as a dissipative soliton. Comparing with the operation of the conventional soliton fiber lasers, where sub-picosecond optical pulses could be routinely generated but the pulse energy is limited at the level of tens of pico-joules, the operation of a dissipative soliton fiber laser can naturally generate large energy strongly chirped optical pulses . Due to its strongly chirped feature, a dissipative soliton can be easily amplified and compressed \cite{no10,no11}. Consequently, large energy ultrashort pulses can be generated. However, because a dissipative soliton fiber laser is operating in the large positive cavity dispersion regime, self-started mode locking in the laser is much more difficult to achieve than in the conventional soliton fiber lasers.

\subsection{Graphene Mode-Locked Fiber Laser}

Unlike the conventional semiconductor saturable absorbers, the energy band diagram of graphene has zero bandgap and a linear dispersion relation \cite{no12}. These unique energy band properties combined with the Pauli blocking principle renders graphene a full band ultrafast saturable absorber \cite{no13}. Fig. 1a shows a schematic illustration of the energy band structure and photon absorption of graphene. Li et al \cite{no14}] had experimentally shown that atomic layer graphene could absorb a considerable amount of infrared light without any bandwidth limitation. In previous papers we have also experimentally shown passive mode locking of an erbium-doped fiber laser \cite{no15)} and a solid-state Nd:YAG ceramic laser \cite{no16} respectively using graphene as a mode locker. It is to point out that determined by its unique energy band structure the saturation intensity of graphene is wavelength sensitive. We had experimentally measured the saturation intensity of graphene at ~1.06 ${\mu}$m and ~1.56 ${\mu}$m, respectively.  It varies from ~0.87 MW* cm$^{-2}$ at 1.06?m to ~0.71 MW*cm$^{-2}$ at 1.56 ${\mu}$m \cite{no13,no15,no16}.

\begin{figure}
  \centering
  \subfigure[]{
    \label{fig:subfig:a} 
    \includegraphics[width=8cm]{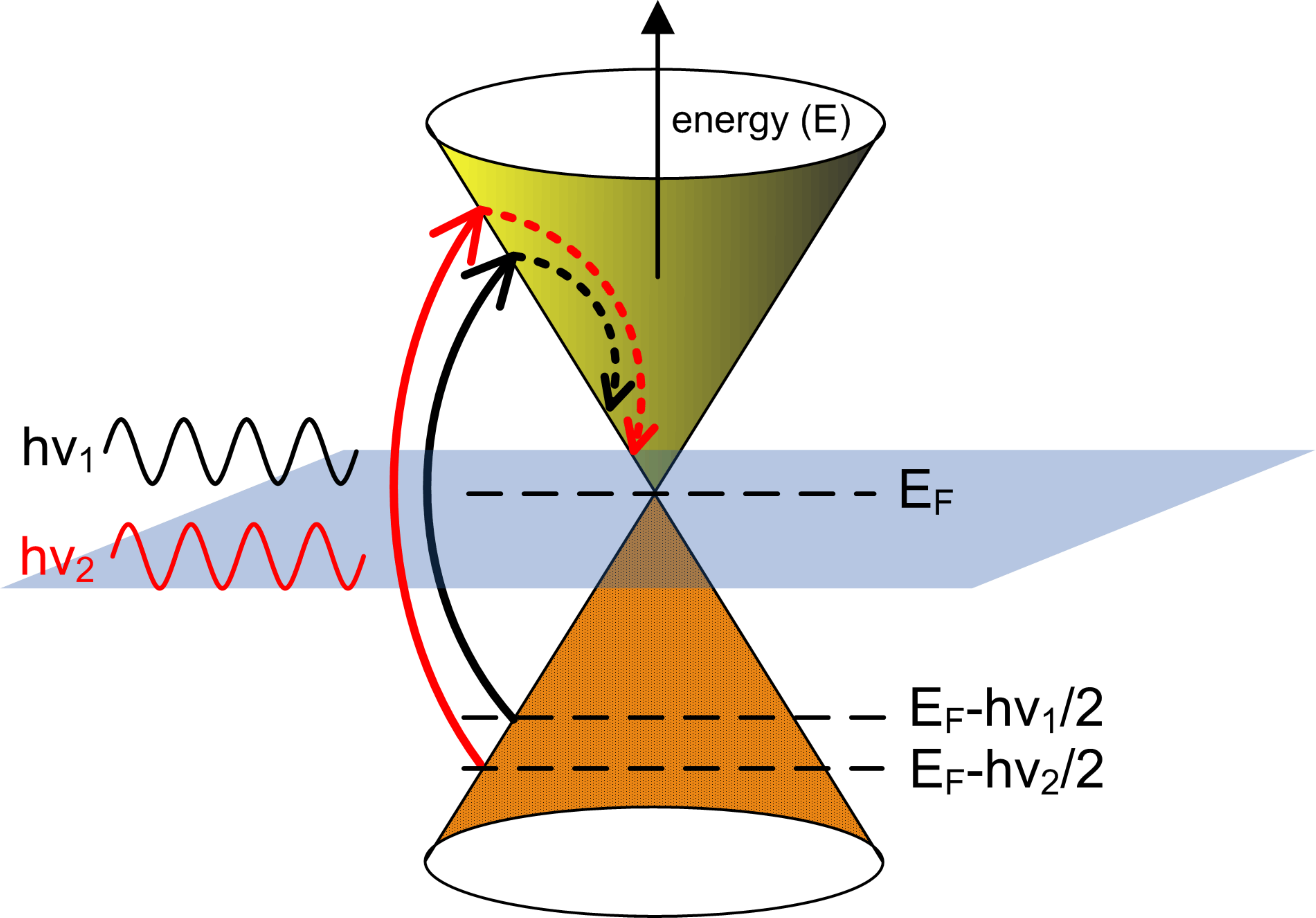}}
  \hspace{1in}
  \subfigure[]{
    \label{fig:subfig:b} 
    \includegraphics[width=8cm]{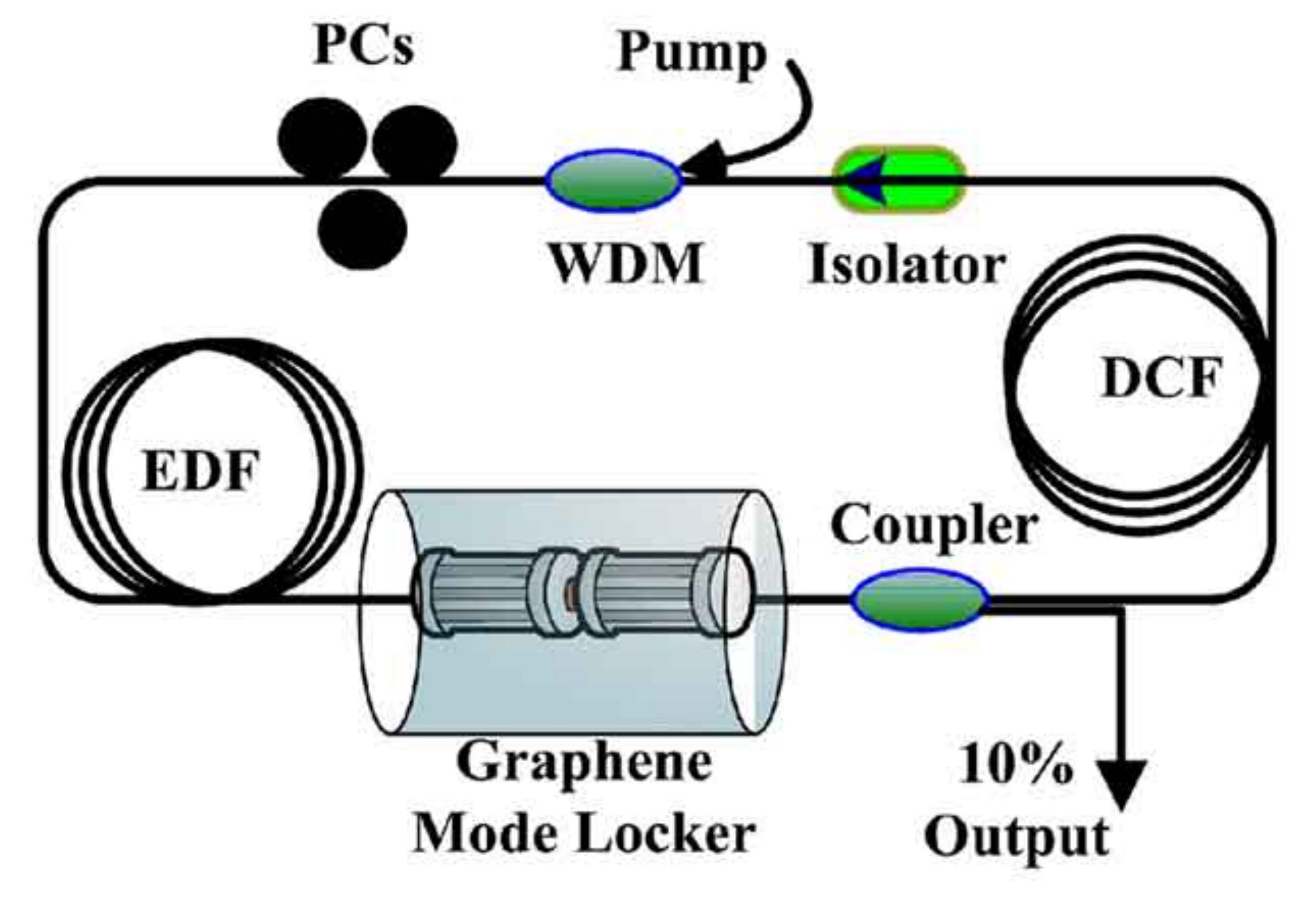}}
  \caption{(a) Schematic of graphene's energy band structure and photon absorption. (b) Schematic of the dissipative soliton fiber laser. WDM denotes: wavelength division multiplexer,
  EDF denotes erbium-doped fiber, PC denotes polarization controller, and DCF denotes dispersion compensation fiber.}
  \label{fig:subfig} 
\end{figure}

\section{Experimental studies}

To exploit the wideband saturable absorption of graphene for wavelength tunable soliton generation, we further specially designed an erbium-doped dissipative soliton fiber laser. We note that dissipative soliton operation of fiber lasers mode locked with SWCNTs was reported by Kieu and Wise \cite{no7}, and Z. Sun et al.  However, dissipative soliton operation of fiber lasers mode locked with atomic layer graphene has not been demonstrated. Moreover, wavelength tunable dissipative soliton lasers were never reported. Our fiber laser is schematically shown in Fig. 1b.  The laser cavity is made of a piece of 5.0 m erbium-doped fiber (EDF) with erbium-doping concentration of 2280 ppm and a group velocity dispersion (GVD) parameter of -32 (ps/nm)/km, a total length of 9.0m single mode fiber (SMF) with GVD parameter of 18 (ps/nm)/km from optical components, and 118m dispersion compensation fiber (DCF) with GVD parameter of about -2 (ps/nm)/km. A polarization independent isolator was used to force the unidirectional operation of the ring, and an intra-cavity polarization controller (PC) was used to adjust the cavity birefringence. The laser was pumped by a high power Fiber Raman Laser of wavelength 1480 nm. The graphene saturable absorber was inserted in the cavity through transferring a piece of free standing few layers graphene film onto the end facet of a fiber pigtail via the Van der Walls force. Details on the graphene preparation and graphene saturable absorber characterization were reported in \cite{no13}. The main difference of the current fiber laser from that reported in \cite{no15} is that normal dispersion EDF instead of anomalous dispersion EDF was employed as the gain medium. To assure large net positive cavity dispersion we also added a long piece of DCF to the cavity. In addition, we deliberately introduced large cavity birefringence in the laser, which strengthens the effect of the artificial birefringence filter formed in the cavity \cite{no17,no18}.

Due to the saturable absorption of graphene, mode locking of the laser was always achieved. As the total cavity dispersion is ~0.3047ps$^{2}$, the mode locked pulse was shaped into a dissipative soliton. Fig. 2 shows a typical dissipative soliton operation state of the laser. Fig. 2a is the optical spectrum of the soliton. It has the characteristic sharp steep spectral edges of the dissipative solitons \cite{no8,no9,no10}. The 3dB spectral bandwidth is ~7.2 nm. Using a 50 GHz high-speed oscilloscope (Tektronix CSA 8000) together with a 45 GHz photo-detector (New Focus 1014), we measured the pulse profile of the dissipative solitons, as shown in Fig. 2b. The pulse width is ~49 ps. The time-bandwidth product of the pulses is ~44.3, indicating that they are strongly chirped. Fig. 2c shows the oscilloscope trace of the mode locked pulse train. The pulse-to-pulse separation is ~660 ns, which matches the cavity roundtrip time. In our experiment we had also measured the mode locked pulses with a commercial autocorrelator and confirmed that no fine structures within the pulses.

\begin{figure}
  \centering
  \subfigure[]{
    \label{fig:subfig:a} 
    \includegraphics[width=7.5cm]{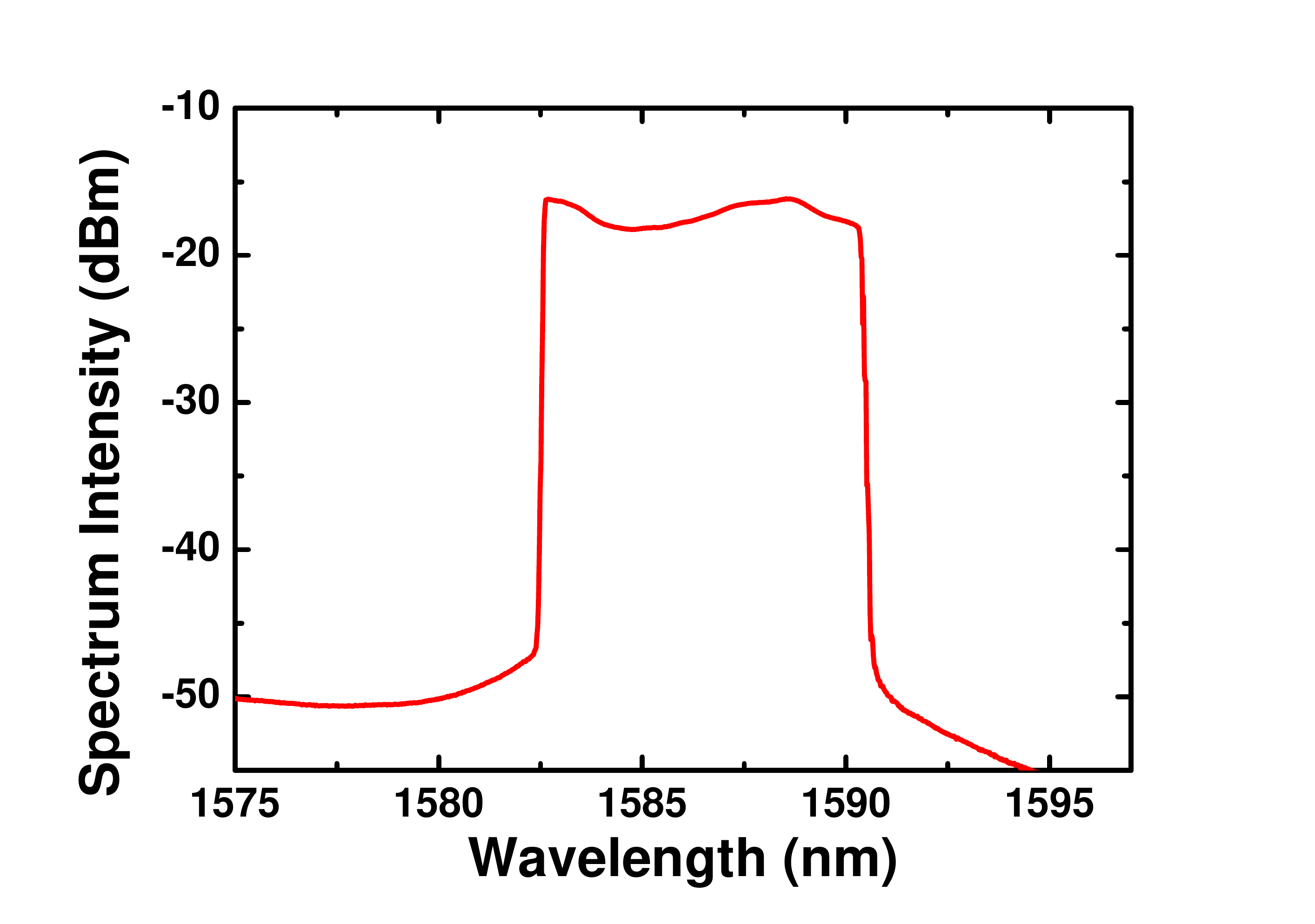}}
  \hspace{1in}
  \subfigure[]{
    \label{fig:subfig:b} 
    \includegraphics[width=7.5cm]{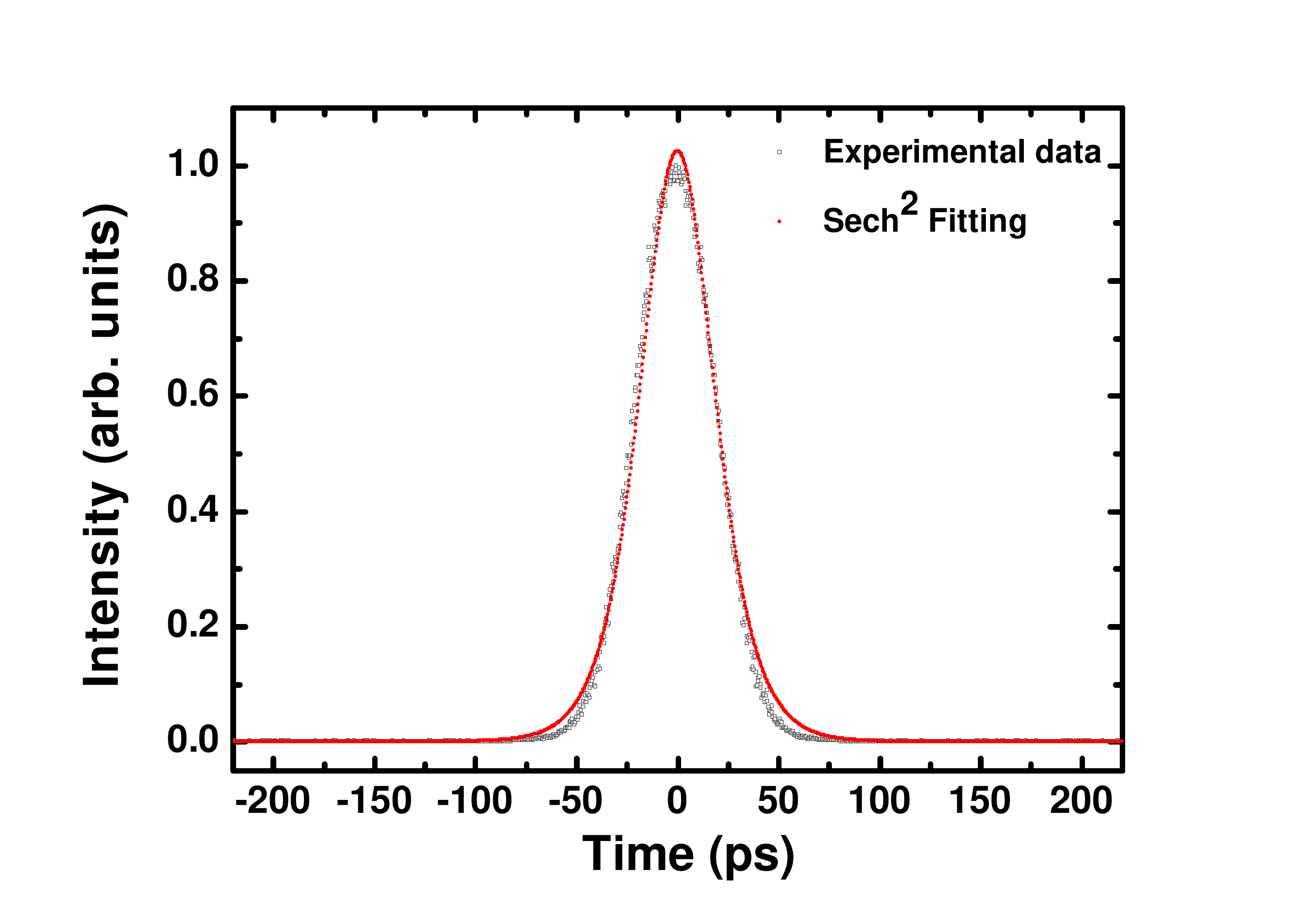}}
      \subfigure[]{
    \label{fig:subfig:b} 
    \includegraphics[width=7.5cm]{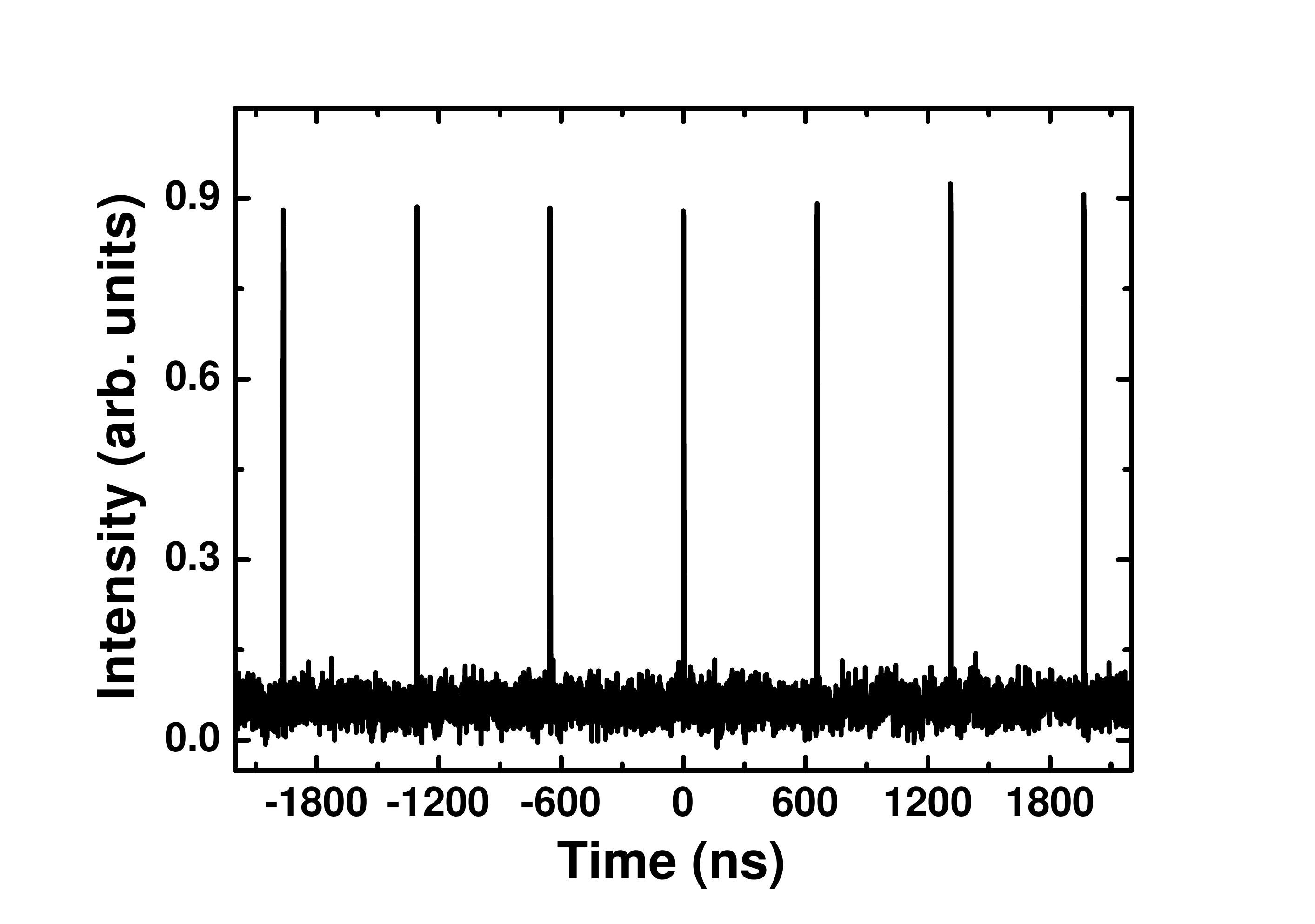}}
  \caption{Dissipative soliton operation of the laser. (a) Optical spectrum measured. (b) Pulse profile. (C) Oscilloscope trace of pulse train.}
  \label{fig:subfig} 
\end{figure}

In weakly birefringent cavity fiber lasers, where the cavity birefringence induced artificial birefringence filter has a large bandwidth; the filtering effect of the birefringence filter could normally be ignored. In our laser, the effect of cavity artificial birefringence filter no longer can be ignored, instead it together with the erbium fiber determines the effective laser gain and gain profile. As transmission of the artificial birefringence filter varies with the cavity birefringence, therefore, in our laser simply adjusting the orientation of the intra cavity polarization controller both the peak wavelength and the pulse width of the formed dissipative solitons could be continuously tuned. Fig. 3a shows the optical spectral evolution of the formed dissipative solitons with the orientation variation of the intra cavity polarization controller. It is to see that the wavelength of the formed dissipative solitons could continuously be shifted from 1570nm to 1600nm. We note that in order to keep stable dissipative soliton operation, as the PC orientation changed, the pumping strength was also adjusted. Fig. 3b further shows the soliton pulse width and the 3dB spectral bandwidth variations. The soliton pulse width changed from ~140ps to ~40ps, and the corresponding 3dB spectral bandwidth changed from ~3nm to ~9nm. It is worth of mentioning that in our experiment we had also constructed a wavelength tunable conventional soliton fiber laser mode locked with graphene. Continuous wavelength tuning of the formed solitons in the same wavelength range (1570 nm -1600 nm) was also observed.

\begin{figure}
  \centering
  \subfigure[]{
    \label{fig:subfig:a} 
    \includegraphics[width=8cm]{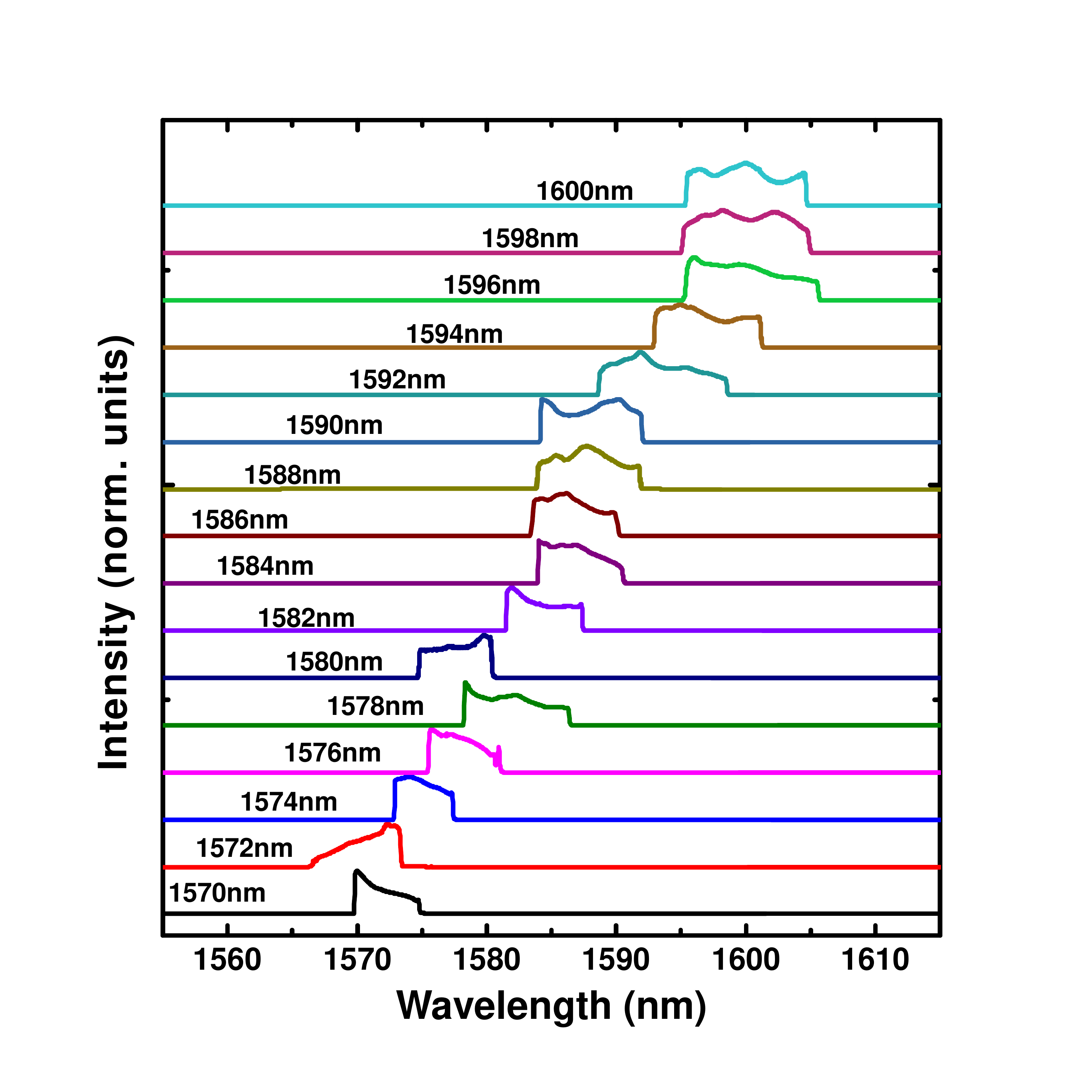}}
  \hspace{1in}
  \subfigure[]{
    \label{fig:subfig:b} 
    \includegraphics[width=8cm]{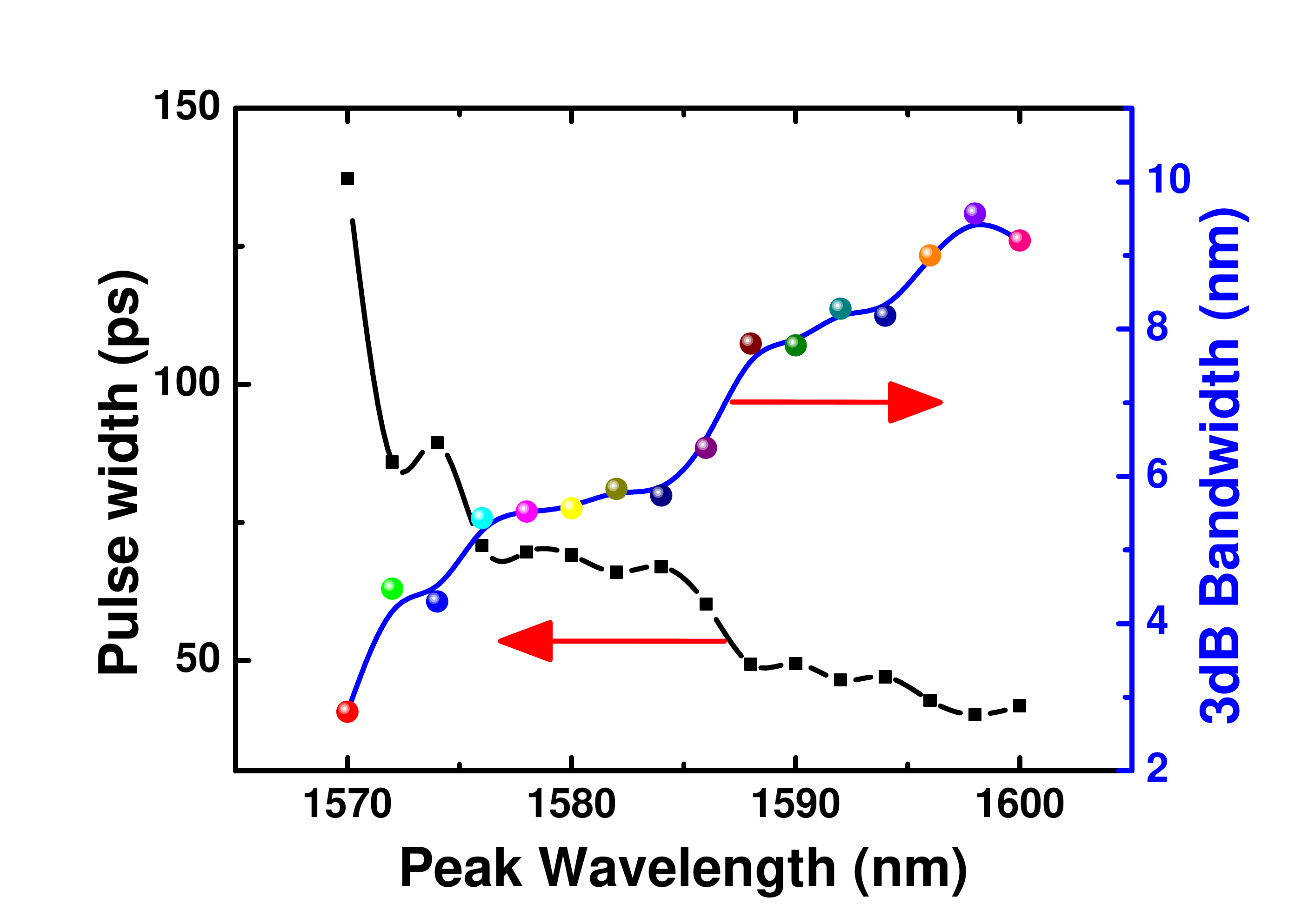}}
  \caption{(a) Wideband output spectra from 1570 nm to 
  1600 nm and inserted numbers indicates the corresponding peak wavelengths; (b) Output pulse duration and spectral bandwidth at different peak wavelengths.}
  \label{fig:subfig} 
\end{figure}

We point out that the artificial cavity birefringence filter in our laser has played the same role as the intra-cavity tunable band-pass filter reported in \cite{no2}. Although with an intracavity tunable band-pass filter even larger wavelength tuning range may be achievable, our laser has nevertheless a much simple configuration and less demand on the cavity components. In addition, in contrast to using tunable filter with only 3 nm bandwidth \cite{no2)}, the current artificial birefringence filter possesses the advantage of always having a relatively broader transmission bandwidth variable through adjusting the cavity birefringence. The effective gain bandwidth limitation effect in our laser is less significant. Consequently, large energy dissipative soliton pulses could be formed in the laser. In our laser dissipative solitons with single pulse energy of 2.3 nJ have been directly generated. It is anticipated that larger pulse energy could be readily generated through further improving the cavity design, such as larger cavity output ratio. Wavelength tunable lasers have widespread applications in many fields, e.g. laser spectroscopy, biomedical research and telecommunications. The developed wavelength tunable dissipative soliton fiber laser provides a cost effective solution for such a light source. Furthermore, the fiber laser could also be used as a compact, cost effective seed source for the generation of large energy wavelength tunable ultrashort pulses.

\section{Conclusion}

In conclusion, we have reported a wavelength tunable erbium-doped dissipative soliton fiber laser with atomic layer graphene as the mode locker. We have shown experimentally that by taking advantage of the artificial cavity birefringence filter effect and the broad band saturable absorption of atomic layer graphene, a wide range wavelength tunable dissipative soliton fiber laser could be constructed. Our studies clearly show that graphene could be a promising wide-band saturable absorber.

\section{Acknowledgement}

 The work is funded by the National Research Foundation of Singapore under the Contract No. NRF-G-CRP 2007-01. K. P. Loh wishes to acknowledge funding support from NRF-CRP Graphene Related Materials and Devices (R-143-000-360-281).

\section{Citations and References}\label{sec:endnotes}

\end{document}